\documentclass[aps,prd,reprint,longbibliography,amsmath,amssymb,superscriptaddress,raggedbottom]{revtex4-2}
\usepackage[T1]{fontenc}
\usepackage{lmodern}
\usepackage{graphicx}
\usepackage{bm}
\usepackage{xurl}
\usepackage{microtype}
\usepackage{hyperref}
\usepackage{orcidlink}
\hypersetup{colorlinks=true,linkcolor=blue,citecolor=blue,urlcolor=blue}

\begin{document}

\title{Lightest-mass suppression and weaker Majorana cancellation in Froggatt--Nielsen textures}

\author{Boliang Yu\,\orcidlink{0009-0006-4664-5068}}
\email[Corresponding author: ]{yuboliang@sjtu.edu.cn}
\author{Ruixin Zhou\,\orcidlink{0009-0004-9303-6674}}
\affiliation{School of Physics and Astronomy, Shanghai Jiao Tong University,
800 Dongchuan Road, Shanghai 200240, China}

\begin{abstract}
The effective Majorana mass is the length of a sum of three complex electron-flavour contributions. A mass-generation model determines both their sizes and how they cancel. We compare direct effective coefficients with a three-right-handed-neutrino seesaw measure at fixed Froggatt--Nielsen charges and common lepton conditions. In the normal-ordering benchmark truncated-normal ensemble, seesaw generation lowers the lightest-mass median by a factor of 3.2 while raising the effective-mass median by 13\%. A comparison reweighted to a common lightest-mass distribution locates nearly all of the coherent mean-square increase in weaker destructive solar-pair interference. In a fixed-amplitude phase-exchange control, the response follows the conditional phase law. A second texture exposes the competing spectral effect: smaller amplitudes can outweigh weaker cancellation in the original mean square. The interference information also affects how a prediction can be used. Replacing induced phases by independent uniform phases at unchanged posterior masses and electron weights raises the benchmark direct median by 29\%. The matrix ensemble therefore predicts a joint arrangement of amplitudes and phases whose coherent information is lost when only masses and electron-flavour weights are retained.
\end{abstract}

\maketitle

\section{Introduction}\label{sec:1}

\begin{figure*}[t]

\centering

\includegraphics[width=\textwidth]{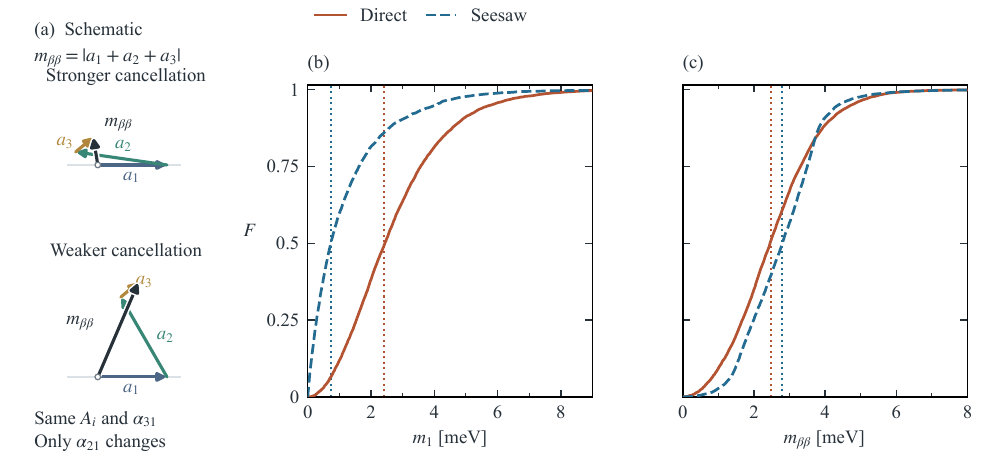}

\caption{Panel (a) illustrates stronger and weaker cancellation at identical $A_i$ and $\alpha_{31}$; only $\alpha_{21}$ changes. Both sketches share a length scale. Panels (b,c) show empirical NO cumulative distributions of $m_1$ and $m_{\beta\beta}$ for the benchmark TN ensembles. Here $F(x)=P(X\leq x)$ for the plotted mass $X$. Dotted lines mark population-averaged medians, which need not coincide with the medians of the displayed distributions.}

\label{fig:1}

\end{figure*}

Neutrino oscillations have established a pattern of mass splittings and flavour mixing, but do not determine the lightest neutrino mass or the Majorana phases. These quantities connect theories of lepton flavour to observables beyond oscillations. In particular, neutrinoless double-beta decay probes a coherent combination of neutrino masses and mixing if neutrinos are Majorana particles and light-neutrino exchange supplies the amplitude \cite{ref1,ref2}. A flavour model's prediction for this process must therefore describe how the mass contributions interfere, as well as how large they are.

The Froggatt--Nielsen (FN) mechanism relates fermion hierarchies to flavour charges and powers of a small symmetry-breaking parameter \cite{ref3,ref4}. The charges organize the suppression of matrix entries, while their remaining complex coefficients are undetermined. Statistical descriptions of these coefficients, developed in neutrino anarchy \cite{ref5,ref6} and charge-weighted flavour models \cite{ref7,ref8}, turn a texture into distributions for masses, mixing and phases. Conditioning on lepton data then selects the configurations compatible with the observed spectrum and mixing. The resulting predictions depend both on the charge hierarchy and on how the coefficient ensemble populates those configurations.

Earlier FN studies have obtained few-meV effective masses and destructive Majorana-phase preferences \cite{ref9,ref10}. Direct and seesaw constructions have been compared for mixing and mass-gap observables \cite{ref7} and for absolute-mass distributions \cite{ref11}. Broader charge scans and alternative coefficient prescriptions have also been studied \cite{ref8,ref11,ref12}. We ask how the generating prescription weights interference at given physical contribution sizes, and what this implies for using its mass predictions to reconstruct a coherent observable.

\begin{samepage}
The distinction can be seen in the effective Majorana mass (also denoted $m_{ee}$). In terms of the PMNS mixing matrix $U$,

\nopagebreak[4]

\[
\begin{gathered}
m_{\beta\beta}=|a_1+a_2+a_3|,
\\
 a_i=m_iU_{ei}^2,\\
 |a_i|=A_i=m_i|U_{ei}|^2.
\end{gathered}
\]
\end{samepage}

Masses and electron-flavour mixing determine the vector lengths. The relative phases $\alpha_{21}$ and $\alpha_{31}$ give the directions of $a_2$ and $a_3$ relative to $a_1$. A small effective mass can reflect small contributions, destructive interference, or both. Shortening an opposing contribution can even enlarge the resultant. The same is possible at fixed lengths if the relative directions change (figure~\ref{fig:1}(a)). A lighter spectrum therefore need not imply a smaller coherent amplitude.

We compare independent effective Majorana coefficients with coefficients generated by a seesaw containing three right-handed neutrinos. Within each pair, the charges, hierarchy parameter, overall-scale law and lepton condition are common. The right-handed charge factors cancel exactly in the factorized tree-level seesaw \cite{ref7,ref10}, so the comparison preserves the low-energy FN powers while changing the induced coefficient measure. We begin with the texture of Qiu, Wang and Yanagida \cite{ref9}, called the benchmark texture, and then examine a second literature-motivated texture.

For the benchmark in normal ordering (NO), seesaw generation gives a substantially lighter first state but a slightly larger effective Majorana mass (figure~\ref{fig:1}(b,c)). After matching the lightest-mass distributions, we trace the coherent mean-square increase to weaker destructive interference within the first two electron-flavour contributions, the solar pair. Comparing phase laws at unchanged contribution lengths shows that this interference information can carry the response. The second texture displays the competing effect of smaller amplitudes: their reduction can outweigh weaker cancellation in the mean square. Together, these results distinguish the changes in the spectrum from those in its coherent arrangement.

This distinction matters when a flavour-model mass distribution is used in a double-beta-decay calculation. Equation (5.6) of ref. \cite{ref11}, for example, combines mass information with flat integrations over Majorana phases. A phase-independent oscillation likelihood does not guarantee that this reconstruction preserves the conditional phase law of a matrix-generated ensemble. We quantify the resulting distribution change while retaining the posterior masses and electron-flavour weights. Section~\ref{sec:2.1} and appendix~\ref{app:B.2} specify the common-parameter comparison and the reference's Gaussian coefficient law.

Section~\ref{sec:2} defines the two generating prescriptions and their common lepton conditions. Section~\ref{sec:3} presents the spectral response, section~\ref{sec:4} follows its interplay with cancellation, and section~\ref{sec:5} examines the information needed for effective-mass reconstruction and subsequent decay inference. The typical effective masses lie below the tens-of-meV reach planned for searches such as LEGEND-1000 \cite{ref13,ref14}; interpreting their distributions requires distinguishing an upper limit from a positive light-Majorana-exchange signal.

\section{Two textures under a common generator comparison}\label{sec:2}

\subsection{Matrices, charges and physical observables}\label{sec:2.1}

The FN charges determine which flavour directions are suppressed when the flavour symmetry is broken. In the factorized textures considered here, each lepton field contributes a power of the small hierarchy parameter $\lambda$. For the lepton-doublet charges $q_{L_i}$ and charged-singlet charges $q_{e_i}$, we collect these factors into

\[
\begin{aligned}
D_L&=\operatorname{diag}(\lambda^{q_{L_1}},\lambda^{q_{L_2}},\lambda^{q_{L_3}}),\\
D_e&=\operatorname{diag}(\lambda^{q_{e_1}},\lambda^{q_{e_2}},\lambda^{q_{e_3}}).
\end{aligned}
\tag{1}\label{eq:1}
\]

The charged-lepton Yukawa matrix and the light Majorana mass matrix then have the form

\[
\begin{gathered}
Y_e=D_L\widehat Y_eD_e,\\
M_\nu=m_0D_L\widehat C_5D_L,\\
\widehat C_5=\widehat C_5^T.
\end{gathered}
\tag{2}\label{eq:2}
\]

Each light-neutrino entry carries two doublet suppression factors, whereas a charged-lepton entry carries a doublet and a singlet factor. Hats contain the remaining dimensionless complex coefficients, and the positive $m_0$ sets the overall neutrino mass scale. The matrix $D_L\widehat C_5D_L$ gives the dimensionless flavour structure of the Weinberg operator \cite{ref15}, with its common mass scale absorbed into $m_0$. We work in the FN charge eigenbasis with canonical kinetic terms.

The two generating prescriptions differ in the origin of $\widehat C_5$. The direct model assigns independent distributions to its six free complex entries. The seesaw model obtains it from a Dirac Yukawa matrix $\widehat Y_\nu$ and a complex symmetric right-handed mass matrix $\widehat M_R$:

\[
\begin{gathered}
\widehat C_5^{\rm SS}=\widehat Y_\nu\widehat M_R^{-1}\widehat Y_\nu^T,
\\
 \widehat C_5=\begin{cases}
\widehat C_5^{\rm dir},&\text{direct},\\
\widehat C_5^{\rm SS},&\text{seesaw}.
\end{cases}
\end{gathered}
\tag{3}\label{eq:3}
\]

The right-handed FN factors enter both microscopic matrices: $Y_\nu=D_L\widehat Y_\nu D_N$ and $M_R=M_0D_N\widehat M_RD_N$, where $D_N$ contains the right-handed charge suppressions. Tree-level type-I matching \cite{ref16,ref17} gives $M_\nu=-(v_H^2/M_0)D_L\widehat C_5^{\rm SS}D_L$, where $v_H=\langle H^0\rangle$. The factors $D_N$ cancel exactly when $\widehat M_R$ is invertible \cite{ref7,ref10}. With exactly three right-handed states, the two prescriptions therefore share the same low-energy FN powers while differing in the correlations among effective coefficients. Both prescriptions use the same law for the positive mass scale $m_0$; in the seesaw parametrization $M_0=v_H^2/m_0$. The overall seesaw minus sign is absorbed into the common Majorana-matrix phase convention.

The electron flavour probed by double-beta decay is fixed by diagonalizing the charged leptons. Their rotation is therefore part of the coherent prediction. With $U_e^\dagger Y_eY_e^\dagger U_e$ diagonal and ascending,

\[
\begin{aligned}
M_\nu^{(e)}&=U_e^TM_\nu U_e
\\
&=U^*\operatorname{diag}(m_1,m_2,m_3)U^\dagger,\\
m_{\beta\beta}&=|(M_\nu^{(e)})_{11}|=
\left|\sum_i m_iU_{ei}^2\right|.
\end{aligned}
\tag{4}\label{eq:4}
\]

The smaller adjacent squared singular-value gap defines the solar pair. A lower pair is normal ordering (NO), an upper pair inverted ordering (IO); physical labels are assigned before computing oscillation quantities. Appendix~\ref{app:A.1} specifies the ordering condition and its relation to a linear-gap classification.

An unequal doublet charge assignment suppresses the flavour directions by different amounts. The benchmark of ref. \cite{ref9} uses $q_L=(1,\tfrac12,0)$, $q_e=(4,1,0)$ and $\lambda=0.171$, giving $D_L=\operatorname{diag}(\lambda,\lambda^{1/2},1)$ and $D_e=\operatorname{diag}(\lambda^4,\lambda,1)$. For this benchmark and generic nonzero leading minors of order one, the light singular values scale as $(\lambda^2,\lambda,1)$. Meanwhile $(U_e)_{i1}\sim\lambda^{q_{L_1}-q_{L_i}}$, so each contribution to $(U_e^TD_L\widehat C_5D_LU_e)_{11}$ carries $\lambda^{2q_{L_1}}$. Thus

\[
\begin{gathered}
\frac{m_1}{m_3}\sim\lambda^2,
\\
 \frac{m_{\beta\beta}}{m_3}\sim\lambda^2.
\end{gathered}
\tag{5}\label{eq:5}
\]

The same leading charge power enters both observables. These generic scaling estimates organize their scale but do not fix the coefficient-dependent relative spectrum or cancellation. A field-coordinate change transforms the couplings, charges and measure together; the interference products below are rephasing invariant (appendix~\ref{app:A.2}).

The second texture tests the comparison when the two less suppressed doublet directions have equal charges. It has relative charges $q_L=(1,0,0)$ and $q_e=(5,2,0)$ from ref. \cite{ref11}, a common hierarchy parameter $\epsilon=0.27$, and Gaussian coefficients in both sectors. Subtracting four from the reference's left charges $(5,4,4)$ removes common powers while preserving the charged mass ratios, left rotation and dimensionless light texture. The common $\epsilon$ is the midpoint of the reference's separate direct and seesaw examples, defining a matched comparison rather than reconstructing its published posterior. Appendix~\ref{app:C.2} reports the amplitude and interference results for this pair.

\subsection{Coefficient laws and lepton data}\label{sec:2.2}

The coefficient probability law determines how the unknown complex couplings populate spectra and cancellation patterns. In the benchmark baseline, each free coefficient has an independent uniform phase and a modulus drawn from $N(\mu_r,\sigma_r^2)$ conditioned on positivity (TN). The parent normal has location $\mu_r=1$ and width $\sigma_r=0.30$; its truncated mean and variance are recorded with the coefficient-law comparisons in the repository. This applies to $\widehat Y_e$ and to either $\widehat C_5^{\rm dir}$ or $(\widehat Y_\nu,\widehat M_R)$ before forming the product. A Gaussian-neutrino comparison changes only the neutrino law and retains charged TN coefficients; the second texture uses Gaussian coefficients in both sectors. The Gaussian matrix laws and their symmetric-entry variances are specified in appendix~\ref{app:B.2} \cite{ref8,ref11,ref18}.

Both generators use the same independent overall mass scale,

\[
\begin{gathered}
p(m_0)=\frac{1}{m_0\ln(m_{0,\max}/m_{0,\min})},\\
m_{0,\min}\leq m_0\leq m_{0,\max}.
\end{gathered}
\tag{6}\label{eq:6}
\]

Here $m_{0,\min}$ and $m_{0,\max}$ are the lower and upper limits of the overall mass scale; we use $m_{0,\min}=10^{-3}$ eV and $m_{0,\max}=1$ eV. The logarithm normalizes the density over this interval. The scale multiplies the original coefficient matrix, retaining its amplitude as well as its spectral shape. Appendices~\ref{app:A.2} and~\ref{app:C.1} give the scale--shape relation and the alternative coefficient laws.

The inherited $\lambda=0.171$ minimizes the ensemble-dispersion objective in ref. \cite{ref9}, eqs. (13) and (23), which combines quark and lepton observables using ensemble spreads. It is a model-selected calibration point.

We condition the charged-lepton matrices on their two independent mass ratios, $r_e=m_e/m_\tau$ and $r_\mu=m_\mu/m_\tau$. Through $U_e$, this condition also selects the electron direction in the neutrino mass matrix. The reference values are $r_e^{\rm obs}=2.79\times10^{-4}$ and $r_\mu^{\rm obs}=0.0589$ at $10^{12}$ GeV \cite{ref9,ref19}, with

\[
\begin{gathered}
\chi_e^2=\sum_{a=e,\mu}
\left[\frac{\ln(r_a/r_a^{\rm obs})}{\sigma_e}\right]^2,
\\
 L_e=e^{-\chi_e^2/2}.
\end{gathered}
\tag{7}\label{eq:7}
\]

The width $\sigma_e$ is the model tolerance in the natural logarithms of the charged-mass ratios. Its baseline value is $\sigma_e=0.20\ln 10$, equivalent to 0.20 dex; it describes model dispersion rather than experimental mass uncertainty. Changing it changes the selected charged rotations and therefore the electron projection. Appendix~\ref{app:C.1} examines this dependence and neighbouring common hierarchy parameters. The matrices define an effective texture conditioned on low-energy oscillations, using the quoted charged ratios as a broad reference. Appendix~\ref{app:C.3} gives the scale interpretation and the one-loop SM effective-operator correction \cite{ref22}.

Both generators are conditioned on the same NuFIT 6.1 oscillation profiles \cite{ref20,ref21}, using TBoff/base as the baseline. The resulting distributions incorporate the measured mass splittings and mixing information. The adopted profile-based composite likelihood \cite{ref23} retains correlations within each coordinate block; correlations removed by projection are not reconstructed. Appendix~\ref{app:A.1} specifies its construction; the repository provides the six data/profile comparisons.

Masses and low-energy phases are induced jointly by each matrix ensemble under these lepton conditions. No cosmological or double-beta-decay measurement enters the conditioning, so the mass sums and effective masses remain predictions. Appendix~\ref{app:A} gives the probability measure and numerical evaluation.

\section{Different mass responses}\label{sec:3}

\subsection{A lighter first state with a larger coherent sum}\label{sec:3.1}

The original NO predictions in figure~\ref{fig:1} show different responses to the generating prescription. With TN coefficients, seesaw generation lowers the lightest-mass median by a factor of 3.2 while raising the effective-mass median by 13\%. The lightest-mass distribution moves markedly towards zero. The effective-mass distributions overlap strongly, while the seesaw median lies at a slightly larger mass.

The two observables retain different information about the light matrix. The lightest mass records its smallest singular value; the effective Majorana mass records the residual of electron-flavour contributions after cancellation. Their opposite movements therefore call for an explanation of both the spectrum and the interference.

The measured splittings keep the heavier states close to their oscillation scales. A large relative displacement of the small first mass consequently produces only a modest change in the sum,

\[
\begin{aligned}\sum_i m_i&=m_1+\sqrt{m_1^2+\Delta m_{21}^2}
\\
&+\sqrt{m_1^2+\Delta m_{31}^2}.
\end{aligned}
\tag{8}\label{eq:8}
\]

Both ensembles remain near the NO minimum. The opposing lightest-mass and effective-mass responses also occur in the tested coefficient, charged-selection and local hierarchy comparisons (appendix~\ref{app:C.1}). Their sizes depend on the probability model: Gaussian neutrino coefficients reduce the effective-mass enhancement to 5.8\%, with the charged TN law held fixed.

\subsection{The hierarchy of the light spectrum}\label{sec:3.2}

The selected seesaw spectrum changes predominantly in the first state. The seesaw/direct median ratios of $m_2/m_3$ and $m_1m_2/m_3^2$ are 0.971 and 0.298, respectively. The normalized determinant is strongly reduced while the second-to-third mass ratio remains nearly unchanged. The spectral change therefore involves an additional suppression of the lightest state, beyond an enhancement of the heaviest direction.

A Dirac matrix close to rank two provides a route to this pattern. With $\widehat M_R=W\operatorname{diag}(\rho_a)W^T$, write

\[
\begin{aligned}
\widehat C_5^{\rm SS}&=\sum_a\frac{b_ab_a^T}{\rho_a},
\qquad b_a=\widehat Y_\nu W_a^*,\\
|\det\widehat C_5^{\rm SS}|&=
\frac{|\det\widehat Y_\nu|^2}{|\det\widehat M_R|}.
\end{aligned}
\]

Here $W_a$ is column $a$ of $W$. At fixed nonsingular $\widehat M_R$, approaching linear dependence among the three Dirac vectors suppresses the determinant quadratically. Their rank-one sum approaches a matrix of rank at most two, allowing the first mass to fall while the second remains finite. A small right-handed singular value provides a different route: it enhances a leading rank-one term and can reduce both lighter-to-heaviest ratios. The selected spectrum favours an interpretation in terms of the Dirac near-rank-two channel. Consistently, equal-strength reductions of microscopic spectral spread produce a larger response of $m_1/m_3$ on the Dirac side than on the right-handed side at the original posterior weights (appendix~\ref{app:C.1}). The final normalized hierarchy reflects both matrices and their conditional selection; the determinant identity identifies the route, while the controlled deformation tests its spectral sensitivity.

Haba, Shimizu and Yamada find stronger hierarchies and a downward tendency of effective-mass peaks in Gaussian-anarchy product comparisons \cite{ref24}. Their construction varies the number of random factors across Dirac, seesaw and double-seesaw matrices, without FN charges or a direct symmetric Majorana baseline. Here the FN factors, electron projection and common lepton condition select a different ensemble: a more hierarchical light spectrum coexists with a slightly larger effective-mass median.

\section{Conditional interference beyond the mass distribution}\label{sec:4}

The opposite responses in figure~\ref{fig:1} raise a question about cancellation: how much of the effective-mass change accompanies the altered mass spectrum, and what remains in the relative directions? We first compare ensembles with the same lightest-mass distribution. We then compare phase laws at unchanged electron-flavour contribution lengths, which depend on mixing as well as on mass.

\subsection{Weaker cancellation within the solar pair}\label{sec:4.1}

Choose the electron-row phase so that its first contribution is real. The effective mass becomes

\[
\begin{gathered}
m_{\beta\beta}=|A_1+A_2e^{i\alpha_{21}}+A_3e^{i\alpha_{31}}|,
\\
 A_i=m_i|U_{ei}|^2.
\end{gathered}
\tag{9}\label{eq:9}
\]

The squared magnitude separates into the contribution squares and their pairwise interference,

\[
\begin{gathered}
m_{\beta\beta}^2=Q+I_{12}+I_{13}+I_{23},
\\
 Q=\sum_i|a_i|^2,\\
I_{ij}=2\operatorname{Re}(a_ia_j^*).
\end{gathered}
\tag{10}\label{eq:10}
\]

In particular, $I_{12}=2A_1A_2\cos\alpha_{21}$. Opposing solar-pair contributions have $I_{12}<0$ and reduce the resultant. A less negative value leaves a larger coherent mean square. We use the mean square to locate the interference change because these terms add exactly, and the median to describe the size of a typical resultant.

\begin{figure*}[t]

\centering

\includegraphics[width=\textwidth]{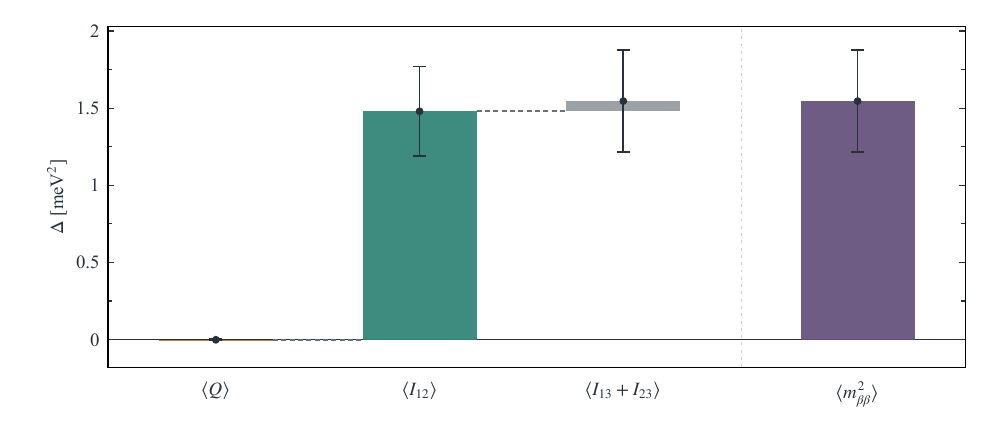}

\caption{Solar-pair interference dominates the seesaw-minus-direct change at a common lightest-mass distribution. The first three bars add the expectation differences for $Q$, $I_{12}$ and $I_{13}+I_{23}$; $\Delta$ is their cumulative sum. The final bar gives the total change in $\langle m_{\beta\beta}^2\rangle$ from zero. Whiskers show numerical 95\% intervals of cumulative endpoints, including covariance. Appendix~\ref{app:A.3} defines the common-mass target.}

\label{fig:2}

\end{figure*}

The original generators populate different lightest-mass ranges. We assign the same probabilities to lightest-mass intervals in their overlapping range, retaining each ensemble's within-interval structure (appendix~\ref{app:A.3}). The effective-mass median remains larger in the seesaw ensemble after this matching. To locate the associated interference change, we examine the mean square under these same matched distributions. Figure~\ref{fig:2} shows nearly equal amplitude-square sums and a less negative solar-pair contribution in the seesaw ensemble; the solar-pair change supplies nearly all of the mean-square increase. The mass-resolved data show contributions across the shared range.

Applying independent uniform phases to these same matched configurations, with their masses and electron weights unchanged, removes almost all of their median difference. Both original generators have uniform primitive coefficient phases; matrix generation and the lepton condition nevertheless induce different low-energy cancellations. Matching the lightest mass still leaves differences in the three contribution lengths. Since $I_{12}$ also contains $A_1A_2$, the next question is whether the phase law carries a response when those lengths are held fixed.

\subsection{Phase information at unchanged contribution lengths}\label{sec:4.2}

\begin{figure*}[t]

\centering

\includegraphics[width=\textwidth]{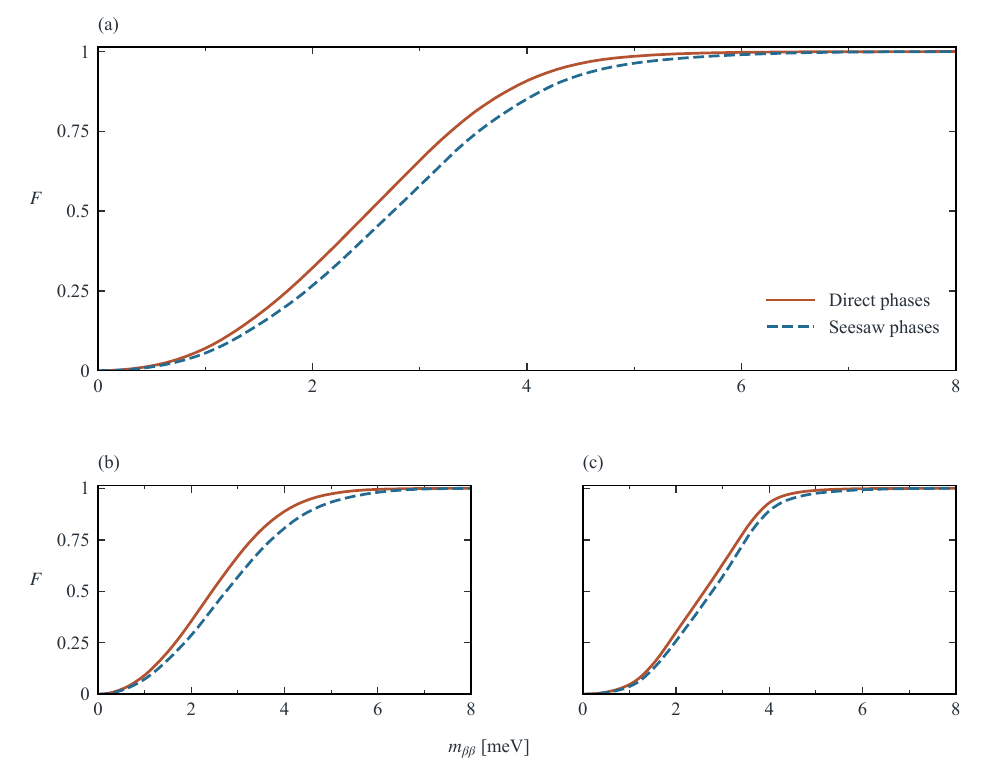}

\caption{Effective-mass reconstructions with direct or seesaw conditional phase pairs. Panel (a) uses common amplitudes; panels (b,c) retain the supported direct and seesaw NO amplitudes, respectively. Each panel holds contribution lengths and weights fixed while exchanging the phase source. Curves show mean cumulative distributions $F$; population-quantile contrasts are provided with the data. Appendix~\ref{app:A.3} defines the support and exchange.}

\label{fig:3}

\end{figure*}

The same contribution lengths can support different coherent predictions. Figure~\ref{fig:3}(a) applies the two conditional phase laws to one common distribution of $A_1,A_2,A_3$. Both laws act on identical lengths with identical weights. Their reconstructed effective-mass distributions differ even though the receiving configurations keep their masses and electron-flavour weights.

Figures~\ref{fig:3}(b,c) repeat the comparison on each generator's own amplitude background within the supported region of its original NO posterior. Within each panel, the background remains fixed and only the paired relative phases are replaced. These backgrounds are distinct from the common lightest-mass-bin comparison of section~\ref{sec:4.1}; appendix~\ref{app:A.3} specifies the supported distributions.

Importing seesaw phases raises the median on the direct amplitude background, while importing direct phases lowers it on the seesaw background. The difference also remains on the common amplitude distribution. Thus a characteristic part of the cancellation information can be carried by the paired conditional phases at given lengths. The exchange keeps the two phases together, preserving their mutual correlation alongside their dependence on the amplitudes.

\subsection{When smaller amplitudes hide weaker cancellation}\label{sec:4.3}

\begin{figure*}[t]

\centering

\includegraphics[width=\textwidth]{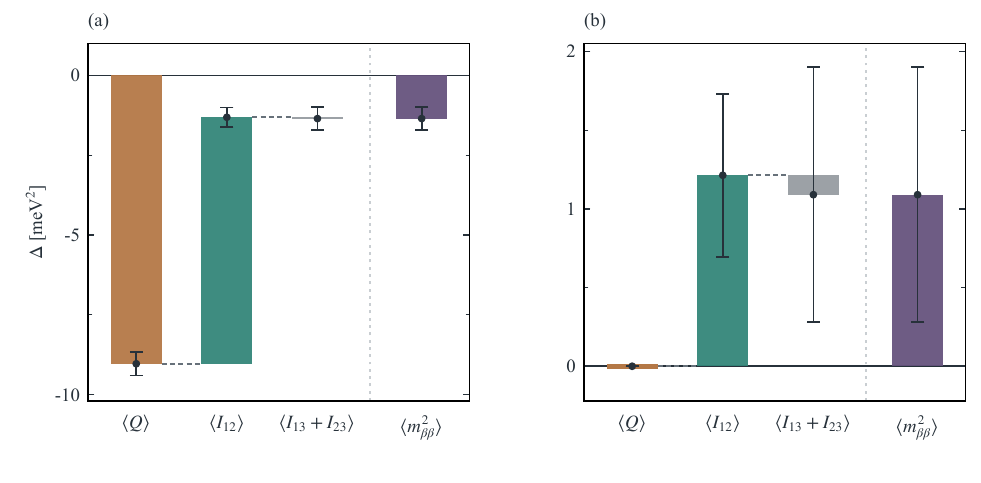}

\caption{Competing amplitude and interference changes in the second texture: (a) original NO posteriors; (b) common amplitudes. As in figure~\ref{fig:2}, $\Delta$ sums the seesaw-minus-direct expectation differences for $Q$, $I_{12}$ and $I_{13}+I_{23}$; the final bar gives the total change in $\langle m_{\beta\beta}^2\rangle$. Whiskers show numerical 95\% intervals of cumulative endpoints, including covariance. The panels have different comparison backgrounds (appendices~\ref{app:A.3} and~\ref{app:C.2}).}

\label{fig:4}

\end{figure*}

The fixed-length comparisons isolate the effect of phase orientations. In an original model prediction, the contribution lengths change as well, so the two effects can compete. We now return to the original NO posteriors of the second texture. Seesaw generation lowers the lightest-mass median and slightly raises the effective-mass median, yet its full mean square decreases. Figure~\ref{fig:4}(a) shows why: smaller contribution squares lower the result, while less negative solar-pair interference raises it. The amplitude decrease is larger, outweighing the weaker cancellation in the original mean square.

Holding the second texture's amplitudes unchanged removes the amplitude-square difference. The phase reconstruction then gives a positive mean-square change, chiefly from the solar pair (figure~\ref{fig:4}(b)). Its median transfer remains unresolved within the numerical interval, while the mean-square and solar-interference changes are resolved (appendix~\ref{app:C.2}). A median increase and a mean-square decrease can therefore coexist when the full distributions change shape. The effective mass is set by the balance of contribution sizes and cancellation in the distribution being considered.

\subsection{Phase preferences at a fixed physical spectrum}\label{sec:4.4}

\begin{figure*}[t]

\centering

\includegraphics[width=\textwidth]{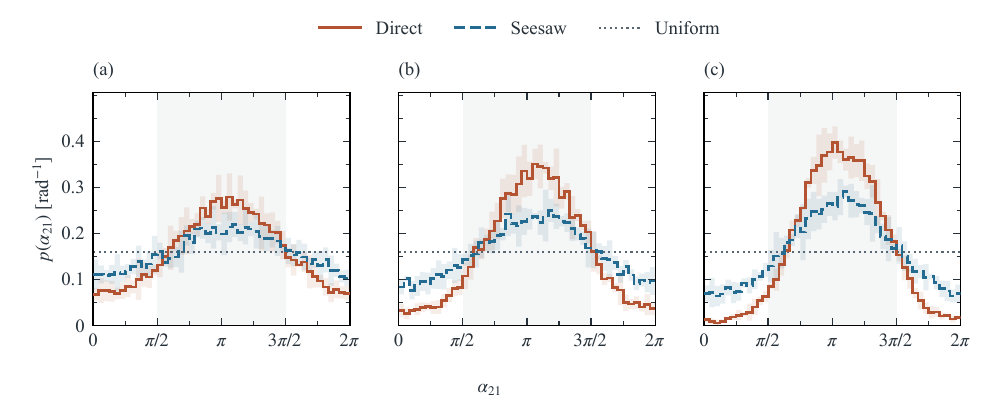}

\caption{Conditional solar-pair phase densities at fixed spectrum and mixing, with $m_1=$ (a) 1, (b) 2 and (c) 4 meV. Shading marks destructive interference, $I_{12}<0$, centred on antiphase at $\pi$; dotted lines denote uniform phase. Steps show the phase bins and bands give pointwise numerical 95\% intervals. Appendix~\ref{app:B.1} specifies the common oscillation inputs and integration.}

\label{fig:5}

\end{figure*}

The phase exchanges establish that the conditional orientations carry predictive information. To see where their preferences enter the matrix model, we now fix the complete light spectrum, mixing angles and Dirac phase. The oscillation weight is unchanged as the Majorana phases vary on these slices. Figure~\ref{fig:5} shows how the two generating measures weight the remaining orientations, with the common charged-lepton condition retained and the remaining microscopic variables integrated. Both favour opposing solar-pair contributions, with stronger opposition in the direct ensemble on the displayed physical slices.

At fixed masses and mixing, changing the relative phases requires different microscopic matrices. Direct generation assigns weights to the effective Majorana entries, which combine mass-weighted contributions. The seesaw representation instead weights the right-handed entries reconstructed from inverse-mass-weighted contributions at a given Dirac matrix. Their coefficient densities continue to distinguish phase orientations even though the oscillation weight is unchanged on a fixed physical slice. Integrating over the remaining microscopic choices gives the preferences in figure~\ref{fig:5}. The matrix expressions in appendix~\ref{app:B.1} show where these phase-dependent weights enter.

\section{Consequences for effective-mass prediction}\label{sec:5}

\subsection{What is lost in flat-phase reconstruction}\label{sec:5.1}

We return to the original NO posteriors to ask what must be retained when a flavour-model prediction is passed to a decay calculation. A matrix ensemble specifies how the electron-flavour contributions cancel as well as how large they are. Retaining only masses and electron weights preserves the lengths and the kinematically allowed effective-mass range \cite{ref2}, but leaves the probabilities within that range unspecified. Independent uniform phases supply one possible replacement law \cite{ref25}. The original ensemble supplies its own phase preferences and their association with the lengths.

\begin{figure*}[t]

\centering

\includegraphics[width=\textwidth]{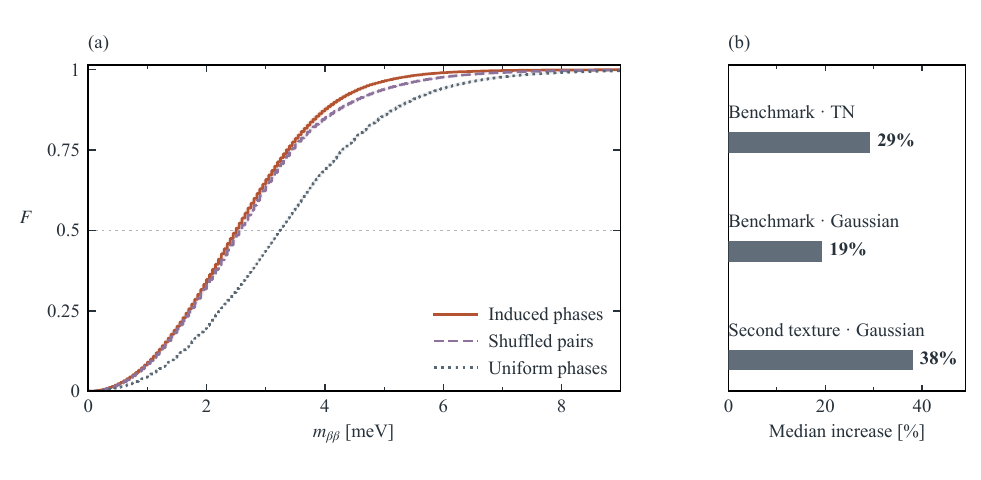}

\caption{Reconstruction after discarding phase information. (a) Benchmark direct TN posterior with induced phases, intact phase pairs shuffled between configurations, or independent uniform phases; masses and electron weights are unchanged. Curves show mean cumulative distributions $F$, with pointwise numerical 95\% bands. (b) Median increase under uniform-phase replacement, from population quantiles. Charged coefficients remain TN for both benchmark neutrino laws and are Gaussian for the second texture. All comparisons condition on NO (appendices~\ref{app:A.3}, \ref{app:C.1} and~\ref{app:C.2}).}

\label{fig:6}

\end{figure*}

Replacing those directions by independent uniform phases shifts the reconstructed effective-mass distribution towards larger values (figure~\ref{fig:6}). In the direct TN benchmark, the median rises by 29\%. The shift is 19\% with Gaussian neutrino coefficients and the charged TN law retained, and 38\% for the second texture with Gaussian coefficients in both sectors. Each comparison keeps the original posterior masses and electron weights unchanged.

The two replacements remove different information. Shuffling intact pairs retains their joint phase population, including the mutual phase correlation, but removes the original association with contribution size. Uniform replacement changes the phase population as well. Their separated curves show that both ingredients matter to the coherent distribution, over a range of effective masses.

For these predictions, a mass-and-weight summary therefore needs the original conditional phase law to reconstruct the same coherent distribution. Alternatively, one can retain the already calculated $m_{\beta\beta}$ distribution for subsequent double-beta-decay inference. Appendix~\ref{app:B.3} expresses this relation in terms of the matrix-induced joint probability law.

\subsection{Central scale, cancellation region and upper tail}\label{sec:5.2}

The few-meV centre reflects the FN suppression in eq.~\eqref{eq:5}, the measured neutrino splittings and the selected electron projection. Relative orientations determine how much of the contribution sum survives. Near cancellation, a small phase change can make a large relative difference to the resultant; near constructive alignment, the same lengths contribute to its upper side. These regions therefore retain information that a median alone does not describe.

The small reactor angle suppresses the heaviest-state electron contribution, $A_3=m_3\sin^2\theta_{13}$, restricting an important route to large effective mass. The released auxiliary comparison examines this selection at a common heavy-mass scale (appendix~\ref{app:C.4}). The tested coefficient measures retain a few-meV centre but populate the rare large-mass configurations very differently (figure~\ref{fig:C1}).

Ordering also matters in the far tail. In IO, both members of the solar pair lie at the atmospheric mass scale and can make substantial electron-flavour contributions. A small posterior IO weight can thus supply much of the remaining large-effective-mass probability. The tail depends on the ordering mixture as well as on the conditional cancellation distribution.

\subsection{What subsequent observations would test}\label{sec:5.3}

To make the effect of a possible observation explicit, consider an ideal effective-mass upper condition $m_{\beta\beta}<t$ or an exceedance $m_{\beta\beta}>t$. Each condition selects part of the existing predictive distribution. The probability assigned to the same event by the two ensembles determines its incremental update of their relative weights.

An upper limit above almost all of both predicted distributions removes little probability from either. Regions where their shapes differ can carry more information about the generating prescription: the deep-cancellation region and the upper side need not have the same relative weights even when the central masses are close. For the second texture, an exceedance of 5 meV reduces the seesaw/direct odds by about a factor of three, even though the seesaw effective-mass median is slightly higher. Appendix~\ref{app:C.4} specifies this NO-conditioned event and its numerical interval; the repository supplies the complete probability updates.

In an experimental analysis, the coherent predictive distribution enters the decay-rate likelihood together with detector response and nuclear matrix elements \cite{ref1,ref26,ref27}. The LEGEND-1000 design studies target tens of meV in effective mass, depending on those matrix elements \cite{ref13,ref14}. An established light-Majorana-exchange signal in that region would test the low-effective-mass expectation shared by the studied ensembles. A null result at that reach would remove little of their probability. Retaining the coherent distribution carries the flavour model's interference information into this later experimental inference.

\section{Conclusions}\label{sec:6}

A mass-generation model determines both the lengths and the relative directions of the electron-flavour contributions to neutrinoless double-beta decay. In the benchmark FN texture, seesaw generation suppresses the first mass state while slightly increasing the effective Majorana mass. A more hierarchical spectrum and a larger coherent resultant coexist because the spectrum does not fix the cancellation.

At a common lightest-mass distribution, the benchmark difference is concentrated in weaker destructive solar-pair interference. The conditional phases also retain a coherent response at unchanged contribution lengths, and their preferences remain different on identical physical spectra. The second texture demonstrates the competing spectral effect: its smaller amplitudes outweigh weaker cancellation in the original mean square. The balance between contribution sizes and interference determines the distribution's response, including the different movements of its median and mean square.

This information has a direct consequence for using a flavour-model prediction. Independent uniform phase replacement at unchanged masses and electron weights raises the benchmark direct TN effective-mass median by 29\%. Its mass posterior alone therefore does not preserve the coherent prediction. Keeping the conditional paired-phase law alongside the amplitudes, or propagating the already calculated effective-mass distribution, retains the information needed for subsequent double-beta-decay inference. FN charges organize the hierarchy; the generating measure also specifies how the surviving contributions interfere.

\textbf{Data and code availability.} The accompanying data and reproducibility package is identified by \href{https://doi.org/10.5281/zenodo.23838671}{Zenodo DOI 10.5281/zenodo.23838671}. It includes joint posterior samples, figure inputs, numerical summaries and the model and inference implementations. Original predictions and controlled reconstructions are identified separately. Repository documentation explains their use, and the plotting scripts reproduce individual numerical panels as PNG files.

\appendix

\setcounter{table}{0}\setcounter{figure}{0}

\renewcommand{\thetable}{A\arabic{table}}

\renewcommand{\thefigure}{A\arabic{figure}}

\renewcommand{\theHtable}{appendix.A.\arabic{table}}

\renewcommand{\theHfigure}{appendix.A.\arabic{figure}}

\section{Probability measures and comparison targets}\label{app:A}

\subsection{Lepton conditions and mass ordering}\label{app:A.1}

For primitive parameters $\vartheta$ and ordering $o$, the probability model is

\[
p(\vartheta\mid D,o)=
\frac{p(\vartheta)\mathbf1_{O(\vartheta)=o}
L_e(\vartheta)L_{{\rm osc},o}(\vartheta)}{Z_o}.
\tag{A.1}\label{eq:A.1}
\]

The primitive coefficient laws and the original mass-scale support are those of section~\ref{sec:2}. The ordering region is part of the matrix model. For ascending singular values, the smaller adjacent squared gap identifies the solar pair before physical mass labels are assigned. The source's linear-gap IO region contains the squared-gap IO region; their disagreement has a squared-gap ratio between $1/4$ and $1/2$, far from the measured hierarchy. Complete classification and evidence checks are provided in the repository.

The composite oscillation condition combines disjoint coordinate blocks,

\[
\begin{gathered}
\chi_{{\rm osc},o}^2=\chi_{\min,o}^2+
\sum_b[P_{b,o}(x_b)-\min P_{b,o}],
\\
 L_{{\rm osc},o}=e^{-\chi_{{\rm osc},o}^2/2}.
\end{gathered}
\tag{A.2}\label{eq:A.2}
\]

The baseline TBoff construction uses the atmospheric block $(\sin^2\theta_{23},\Delta m^2_{3\ell},\delta_{\rm CP})$, the solar block $(\sin^2\theta_{12},\log_{10}(\Delta m^2_{21}/{\rm eV}^2))$, and $\sin^2\theta_{13}$. Here $\ell=1$ in NO and $\ell=2$ in IO. Block minima are removed before the ordering minimum is included once. The six comparisons exchange the TBoff/TByes inputs and the projected solar/reactor blocks. They preserve displayed within-block correlations but do not reconstruct omitted experimental correlations.

Outside a nonperiodic profile grid, we use its nearest-face value plus $\kappa_{\rm ext}d_{\rm grid}^2$, where $d_{\rm grid}$ is distance in grid steps and $\kappa_{\rm ext}=25$ is the chosen continuation strength. This continuation is part of the adopted likelihood, particularly for remote tails. Its definition is separate from numerical integration accuracy.

\subsection{Field coordinates, the original scale and spectral measure}\label{app:A.2}

The physical comparison is invariant under a synchronous right-handed field transformation: $Y_\nu'=Y_\nu W$, $M_R'=W^TM_RW$ gives $Y_\nu'(M_R')^{-1}(Y_\nu')^T=Y_\nu M_R^{-1}Y_\nu^T$. Couplings, charges and probability measures transform together. Likewise $I_{12}=2m_1m_2\operatorname{Re}[U_{e1}^2(U_{e2}^*)^2]$ is invariant under charged-row rephasing and the allowed Majorana column signs. Changing an independent coefficient law changes the probability model.

A matrix-normalized coordinate must retain its correlation with the original scale. For $T(z)=D_L\widehat C_5(z)D_L$, let $a(z)=\|T(z)\|_2$, $\widehat T=T/a$ and $\tau=m_0a$. The original scale endpoints give

\[
\begin{gathered}
M_\nu=\tau\widehat T,\\
p(\tau\mid z)=
\frac{\mathbf1_{m_{0,\min}a(z)\leq\tau\leq m_{0,\max}a(z)}}
{\tau\ln(m_{0,\max}/m_{0,\min})}.
\end{gathered}
\tag{A.3}\label{eq:A.3}
\]

Thus the log-scale law leaves no explicit amplitude power, but its transformed support depends on $a$. Assigning $\tau$ a new independent fixed support would define a different model. More generally, a law $p(m_0)\propto m_0^\alpha$ induces the factor $a^{-(\alpha+1)}\tau^\alpha$ on the transformed support.

For the charged-basis matrix $M_\nu^{(e)}$ of eq.~\eqref{eq:4}, set $H=D_L^{-1}U_e^*M_\nu^{(e)}U_e^\dagger D_L^{-1}$ and $K=\widehat Y_\nu^TH^{-1}\widehat Y_\nu$. A radial coefficient density becomes $f_{ij}(r)/(2\pi r)$ in Cartesian coordinates. The ordered Takagi volume contains

\[
\begin{aligned}
J(m)&=\prod_i m_i\prod_{i<j}|m_i^2-m_j^2|,\\
&J(m)\left|\frac{\partial(m_1,m_2,m_3)}
{\partial(m_{\rm lightest},\Delta m_{21}^2,|\Delta m_{3\ell}^2|)}\right|
\\
&=\frac{m_{\rm lightest}}{4}\Delta m_{21}^2|\Delta m_{3\ell}^2|
\left(|\Delta m_{3\ell}^2|-\Delta m_{21}^2\right).
\end{aligned}
\tag{A.4}\label{eq:A.4}
\]

Here $m_{\rm lightest}=m_1$ in NO and $m_3$ in IO; the index $\ell$ in $\Delta m_{3\ell}^2$ retains its ordering definition in appendix~\ref{app:A.1}. The full Takagi volume is proportional to $J(m)\,d^3m\,d\mu_{\rm Haar}(U)$. Its Haar element is a coordinate volume, not an additional mixing prior.

The direct matrix transformation has real Jacobian $|\det D_L|^{-8}m_0^{-12}$. Integrating the original scale law gives the following coefficient factors. They multiply the common Takagi volume of eq.~\eqref{eq:A.4}, with the charged and Dirac measures still retained and overall constants omitted:

\begin{widetext}
\[
\begin{aligned}
\text{direct:}\quad&|\det D_L|^{-8}\prod_{i\leq j}|H_{ij}|^{-1}\\
&\quad\times\int_{m_{0,\min}}^{m_{0,\max}}
       p(m_0)m_0^{-6}\prod_{i\leq j}f_{ij}(|H_{ij}|/m_0)\,dm_0,\\
\text{seesaw:}\quad&|\det D_L|^{-8}
       |\det\widehat Y_\nu|^8|\det H|^{-8}
       \prod_{i\leq j}|K_{ij}|^{-1}\\
&\quad\times\int_{m_{0,\min}}^{m_{0,\max}}
       p(m_0)m_0^{6}\prod_{i\leq j}f_{ij}(m_0|K_{ij}|)\,dm_0.
\end{aligned}
\tag{A.5}\label{eq:A.5}
\]
\end{widetext}

Here $f_{ij}$ is the radial law of the relevant direct or right-handed entry and allows the entry-dependent Gaussian variance. The common FN factor becomes $\lambda^{-12}$ for the benchmark and $\epsilon^{-8}$ for the second texture. These expressions retain the amplitude and shape weighting of the original model. The full derivation, normalization constants and numerical implementation accompany the data.

\subsection{Common-mass and common-amplitude comparisons}\label{app:A.3}

The original NO ensembles and the controlled comparisons use different probability distributions. For the benchmark common-mass comparison, bin probabilities are first averaged over the independent NO populations and then normalized within the selected mass range, before taking the overlap or restricting occupancy support. Let $p_{g,j}$ denote these range-conditioned bin probabilities. The second texture uses this convention; the Gaussian-neutrino comparison uses full-NO bin probabilities. The shared target is

\[
\begin{gathered}
q_j=\frac{\min(p_{D,j},p_{S,j})}
          {\sum_{k\in\mathcal J}\min(p_{D,k},p_{S,k})},
\\
 j\in\mathcal J.
\end{gathered}
\tag{A.6}\label{eq:A.6}
\]

The benchmark uses supported intervals within $m_1\in[0.25,6]$ meV, at a mass resolution of 0.25 meV. Each generator retains its conditional distribution within an interval. Equal interval probabilities do not fix the full mass spectrum or the contribution lengths. The retained support and weights for every comparison are provided with its data product.

For phase exchange, write $\mathbf A=(A_1,A_2,A_3)$ and $\boldsymbol\phi=(\alpha_{21},\alpha_{31})$. At supported amplitudes from generator $g$, the cumulative prediction using generator $h$'s conditional paired phases is

\[
\begin{aligned}
F_{g\leftarrow h}(y)&=
\int p_g^{\rm sup}(\mathbf A\mid D)\,d\mathbf A
\\
&\quad\int p_h(\boldsymbol\phi\mid\mathbf A,D)\,d\boldsymbol\phi\,
\\
&\qquad\mathbf1_{\{|A_1+A_2e^{i\alpha_{21}}+A_3e^{i\alpha_{31}}|\leq y\}}.
\end{aligned}
\tag{A.7}\label{eq:A.7}
\]

The support is the shared amplitude region and the stated common-mass range. In coordinates $(\ln A_2,\ln(A_1/A_2),\ln(A_3/A_2))$, we intersect the two generators' training marginal intervals, trimming each tail by $\eta_{\rm trim}=0.005$, and restrict the density ratio to $[\rho_{\max}^{-1},\rho_{\max}]$, with $\rho_{\max}=20$. These choices delimit overlap independently of phase or interference sign. The ideal common-amplitude target is proportional to the smaller density on this region. In its empirical construction, each supported sample is weighted by the estimated smaller-density-to-own-density ratio and normalized separately; the two weighted samples are combined equally. Both conditional phase laws act on these identical weighted configurations.

Phases remain paired in the exchange and intact-pair shuffle. Uniform replacement instead draws the two effective phases independently. These operations retain the receiving particle's masses and electron weights; they are controlled reconstructions, not separately refitted matrix models. Density estimation, independent training and evaluation blocks, support arrays and reconstruction checks are documented in the repository.

\subsection{Numerical summaries}\label{app:A.4}

Expectations are evaluated using independent sequential Monte Carlo populations \cite{ref28,ref29}, with separate restricted integrals for rare events \cite{ref30}. Reported mass quantiles average population quantiles. Physical posterior ranges describe the distribution, while numerical 95\% intervals describe the estimated summaries. Common-target contrasts refit their weights under population-block deletion, and total interference errors retain component covariance. The released code specifies all estimators and preserves the independent population identities.

\setcounter{table}{0}\setcounter{figure}{0}

\renewcommand{\thetable}{B\arabic{table}}

\renewcommand{\thefigure}{B\arabic{figure}}

\renewcommand{\theHtable}{appendix.B.\arabic{table}}

\renewcommand{\theHfigure}{appendix.B.\arabic{figure}}

\section{Conditional phase densities}\label{app:B}

\subsection{Coefficient weights at a fixed physical spectrum}\label{app:B.1}

The three slices in figure~\ref{fig:5} use $m_1=1,2,4$ meV. Their common oscillation coordinates were selected before the conditional integrations as the highest-joint-likelihood configuration in the existing benchmark direct populations. The fixed inputs, rounded here, are

\[
\begin{gathered}
(\sin^2\theta_{12},\sin^2\theta_{13},\sin^2\theta_{23})
\\
\simeq(0.311225,\,0.0225319,\,0.467237),\\
\delta_{\rm CP}\simeq-2.71415\ {\rm rad},\\
\Delta m_{21}^2\simeq7.53641\times10^{-5}\ {\rm eV}^2,\\
\Delta m_{31}^2\simeq2.52890\times10^{-3}\ {\rm eV}^2.
\end{gathered}
\]

Both generating measures use these same coordinates at every displayed mass. The heavier masses follow from the fixed splittings. The full-precision inputs and selected source configuration accompany the data.

Let $x$ collect the light masses, mixing angles and Dirac phase. The charged-ratio condition $D_{\rm ch}$ remains imposed, and the original scale, charged rotations and microscopic Dirac variables are still integrated. Integrating the primitive density over these remaining variables gives $W_g(x,\boldsymbol\phi;D_{\rm ch})$ and

\[
p_g(\boldsymbol\phi\mid x,D_{\rm ch})=
\frac{W_g(x,\boldsymbol\phi;D_{\rm ch})}
{\int W_g(x,\boldsymbol\phi';D_{\rm ch})\,d\boldsymbol\phi'}.
\tag{B.1}\label{eq:B.1}
\]

At fixed masses and mixing, changing the relative phases requires different microscopic matrices. Write $U=V\operatorname{diag}(e^{i\alpha_k/2})$ with $\alpha_1=0$, and define $B_0=D_L^{-1}U_e^*V^*$ and $T_D=B_0^{-1}\widehat Y_\nu$. The reconstructed combinations are

\[
\begin{gathered}
H=B_0\operatorname{diag}(m_ke^{-i\alpha_k})B_0^T,\\
K=T_D^T\operatorname{diag}(e^{i\alpha_k}/m_k)T_D.
\end{gathered}
\tag{B.2}\label{eq:B.2}
\]

The effective contribution phases follow from $a_i=m_iU_{ei}^2$ in this fixed mixing convention. Direct generation weights the entries of $H/m_0$; seesaw generation weights the reconstructed right-handed entries $m_0K$ at a given Dirac matrix. The former is mass weighted, the latter inverse-mass weighted. Both depend on relative phases.

At fixed $x$,

\[
|\det H|=|\det D_L|^{-2}m_1m_2m_3.
\tag{B.3}\label{eq:B.3}
\]

This determinant and the Takagi mass-and-angle volume are phase independent. The explicit determinant factor in the seesaw Jacobian cannot itself produce the preference on this slice. The varying factors are the primitive Cartesian coefficient densities and their scale and nuisance integrals. Equations~\eqref{eq:B.2}--\eqref{eq:B.3} locate these weights; the evaluated integrals establish the preference direction in figure~\ref{fig:5}.

\subsection{Gaussian coefficients and unequal left-handed charges}\label{app:B.2}

The Gaussian law for a general complex matrix has real-component variance $\sigma_G^2/2$. A complex symmetric matrix has that diagonal variance and off-diagonal variance $\sigma_G^2/4$, giving the unitary-congruence-invariant density $\pi(G)\propto\exp[-\operatorname{Tr}(G^\dagger G)/\sigma_G^2]$ \cite{ref8,ref11,ref18}. We use $\sigma_G=1$ as the coefficient scale. These are the variance definitions in ref. \cite{ref11}, eqs. (2.24)--(2.25). A common variance rescaling leaves the dimensionless prior quantities in this check unchanged.

At fixed scale and charged rotation, let $b_k$ be the columns of $B_0$. The direct Gaussian weight contains

\[
\begin{aligned}
\operatorname{Tr}(H^\dagger H)&=
\sum_km_k^2\|b_k\|^4+
\\
&\quad 2\sum_{k<l}m_km_l\operatorname{Re}
\left[e^{i(\alpha_k-\alpha_l)}(b_k^\dagger b_l)^2\right].
\end{aligned}
\tag{B.4}\label{eq:B.4}
\]

For equal left charges, $B_0$ is proportional to a unitary matrix and the cross terms vanish. Unequal FN factors remove this orthogonality and allow the coefficient density to depend on relative phases. The Gaussian seesaw law also has unitary congruence invariance in the equal-left control; its induced density need not be Gaussian.

\begin{figure*}[t]

\centering

\includegraphics[width=\textwidth]{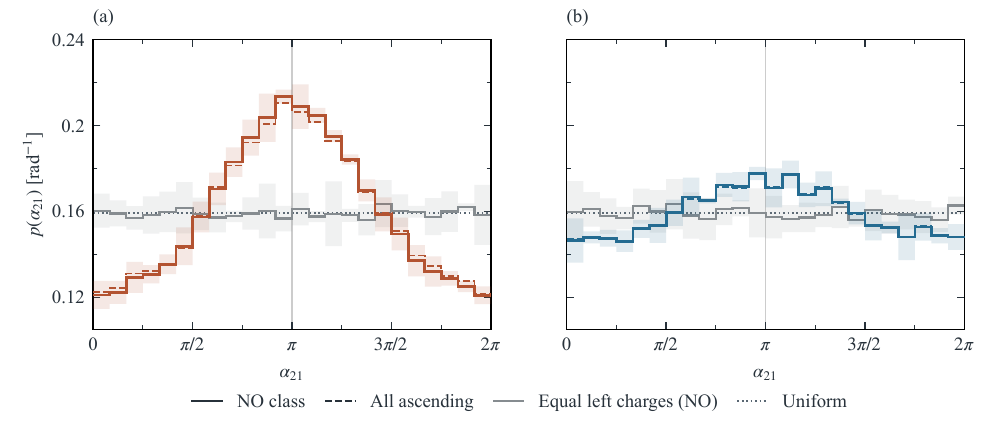}

\caption{Gaussian prior phase densities for (a) direct generation at $\epsilon=0.26$ and (b) three-right-handed-neutrino seesaw at $\epsilon=0.28$. Solid curves select squared-gap NO; dashed curves retain all draws with ascending mass labels. Grey curves show equal-left-charge NO controls; dotted lines denote uniform phase. Bands give pointwise numerical 95\% intervals for NO curves and controls. No lepton likelihood is applied; appendix~\ref{app:B.2} gives the charges and coefficient laws.}

\label{fig:B1}

\end{figure*}

The Gaussian prior examples use relative charges $q_L=(1,0,0)$ and $q_e=(5,2,0)$, with $\epsilon=0.26$ for direct generation and 0.28 for seesaw as in ref. \cite{ref11}. Their equal-left-charge controls set $q_L=(0,0,0)$ while retaining the singlet charges (figure~\ref{fig:B1}). The phase preference is present before imposing the NO selection. These are prior checks, separate from the common-condition second-texture posterior in appendix~\ref{app:C.2}.

\subsection{Joint information in reconstruction}\label{app:B.3}

Let $x$ now include all phase-independent likelihood variables, including charged-lepton ratios. A phase-independent likelihood still reweights a joint matrix-induced law:

\[
p_g(\boldsymbol\phi\mid D)=
\int p_g(\boldsymbol\phi\mid x)\,p_g(x\mid D)\,dx.
\tag{B.5}\label{eq:B.5}
\]

The generating model supplies the conditional phases and the data select the physical configurations on which they act. A mass-and-weight summary preserves contribution lengths and their allowed coherent range, but requires the conditional phase law to preserve the probabilities within that range. Keeping the already computed $m_{\beta\beta}$ distribution is sufficient for subsequent rate inference.

The shuffle and uniform replacements in figure~\ref{fig:6} remove different parts of this information. Their complete distributions, component means and reconstruction estimates accompany the data. The one-harmonic exploratory approximation and its limitations are documented there separately from the density argument.

\setcounter{table}{0}\setcounter{figure}{0}

\renewcommand{\thetable}{C\arabic{table}}

\renewcommand{\thefigure}{C\arabic{figure}}

\renewcommand{\theHtable}{appendix.C.\arabic{table}}

\renewcommand{\theHfigure}{appendix.C.\arabic{figure}}

\section{Model responses and predictive scope}\label{app:C}

\subsection{Coefficient laws and lepton-selection responses}\label{app:C.1}

Changing a coefficient law changes the probability model. Replacing the neutrino TN moduli by a log-normal law matched to their actual radial mean and variance, with charged TN coefficients retained, preserves the generator contrast. Gaussian neutrino coefficients instead reduce the effective-mass enhancement from the TN benchmark's 13\% to 5.8\%, again retaining the charged TN law. The corresponding direct flat-phase reconstruction shifts are 29\% and 19\%. These amplitudes belong to their specified ensembles and are not a single uncertainty interval.

Tightening the charged-lepton condition changes the selected charged rotations and the lightest-mass location more than the effective-mass centre. The released comparisons at equivalent widths $\sigma_e/\ln 10=0.10,0.20,0.30$ dex retain the different mass responses and a positive solar-pair contrast on the common-mass background constructed at each condition. The phase-reconstruction increment also changes with selection. Full distributions, comparison-specific weights and the discrete response summaries accompany the data.

The local hierarchy comparison uses the same parameter for both generators at each point. Its neighbourhood is selected through $\lambda_\pm=\lambda_\star\exp(\pm\sigma_e/p_e)$, where $\lambda_\star$ is the inherited optimum and $p_e=q_{L_1}+q_{e_1}=5$ is the leading electron mass-ratio exponent. This is a local design window set by the adopted charged tolerance, not an experimental or source-fit confidence interval. The three evaluated parameter values retain the different spectral and interference responses. A single common-mass target is used across this local comparison; the complete discrete responses are supplied with the data.

Alternative scale laws proportional to $m_0^{-2}$ or a constant move the lightest-mass medians while retaining the different generator responses and the relative stability of the effective-mass centres. Complete physical ranges, coefficient-law configurations, profile comparisons and numerical validation are supplied as data.

A microscopic spectral control preserves each matrix's singular vectors and geometric mean while reducing its root-mean-square centred logarithmic singular-value spread by the same amount $h$. At $h=0.10$, flattening the Dirac spectrum raises $m_1/m_3$ by a factor 1.454, with numerical 95\% interval $[1.449,1.460]$; the right-handed deformation gives 1.105 with $[1.099,1.110]$. Because the deformation preserves the determinant, it probes spectral shape rather than a change in spectral volume. The original posterior weights are retained without reconditioning the deformed configurations. The full deformation prescription and results accompany the data.

\subsection{Amplitude and interference in the second texture}\label{app:C.2}

Its original effective-mass median rises, while its original mean square falls: the smaller amplitude-square sum outweighs the weaker destructive solar-pair interference (figure~\ref{fig:4}(a)). At common amplitudes, removing this amplitude difference exposes a positive mean-square contrast of $1.089\pm0.810$ meV$^2$, chiefly the solar-pair contrast $1.213\pm0.518$ meV$^2$. The common-amplitude median transfer, $0.141\pm0.172$ meV, remains unresolved. These are numerical 95\% intervals for distinct summaries of the specified reconstruction.

Uniform phases applied at unchanged original masses and electron weights raise the direct median by about 38\%. This result has an all-Gaussian charged sector as well as different charges; comparing it with the benchmark's 19\% Gaussian-neutrino result therefore changes both features. The complete original and controlled distributions are archived separately.

\subsection{Effective-theory scale evolution}\label{app:C.3}

Below all heavy thresholds, one-loop SM running has the charged-basis form \cite{ref22}

\[
\begin{aligned}
M_\nu^{(e)}(\mu)&=I_0\mathcal D\,M_\nu^{(e)}(\Lambda)\mathcal D,\\
\mathcal D&=\operatorname{diag}(D_e^{\rm RG},D_\mu^{\rm RG},D_\tau^{\rm RG}),\\
D_a^{\rm RG}&=\exp\left[\frac{3}{32\pi^2}
\int_{\ln\mu}^{\ln\Lambda}y_a^2(t)\,dt\right].
\end{aligned}
\tag{C.1}\label{eq:C.1}
\]

The superscript distinguishes running factors from the FN charge matrix. Extracting the common factor $I_0(D_e^{\rm RG})^2$ gives $m_{0,\rm eff}=I_0(D_e^{\rm RG})^2m_{0,\rm high}$. The remaining flavour factor is $\operatorname{diag}(1,D_\mu^{\rm RG}/D_e^{\rm RG},D_\tau^{\rm RG}/D_e^{\rm RG})$. Its unit electron entry leaves $m_{\beta\beta}$ unchanged pointwise at fixed $m_{0,\rm eff}$ and matrix shape. Masses and mixing still move, so their lepton-condition weights and the conditional effective-mass distribution can change.

We retain eq.~\eqref{eq:6} for $m_{0,\rm eff}$; its equivalent high-scale support is divided by the extracted common factor. The finite check assumes the SM effective theory from $10^{12}$ to $10^2$ GeV with constant $y_\tau=0.01$, giving non-universal corrections of order $10^{-5}$. In the baseline TN/TBoff/base comparison, reconditioning preserves the primary mass and interference responses within their numerical resolution. Separate endpoint moves have negligible effects in that tested range. Full response arrays and quadrature comparisons accompany the data.

The calculation does not include neutrino-Yukawa running above or between heavy thresholds, where successive matching is required \cite{ref32}. Its scope is the effective-texture comparison, rather than arbitrary ultraviolet completions.

\subsection{High effective masses and ideal event conditions}\label{app:C.4}

For a specified measure, let $S_o(t)=P(m_{\beta\beta}>t\mid D,o)$. Ordering-mixed predictions use

\[
P(m_{\beta\beta}>t\mid D)=
\frac{Z_{\rm NO}S_{\rm NO}(t)+Z_{\rm IO}S_{\rm IO}(t)}
{Z_{\rm NO}+Z_{\rm IO}}.
\tag{C.2}\label{eq:C.2}
\]

\begin{figure*}[t]

\centering

\includegraphics[width=\textwidth]{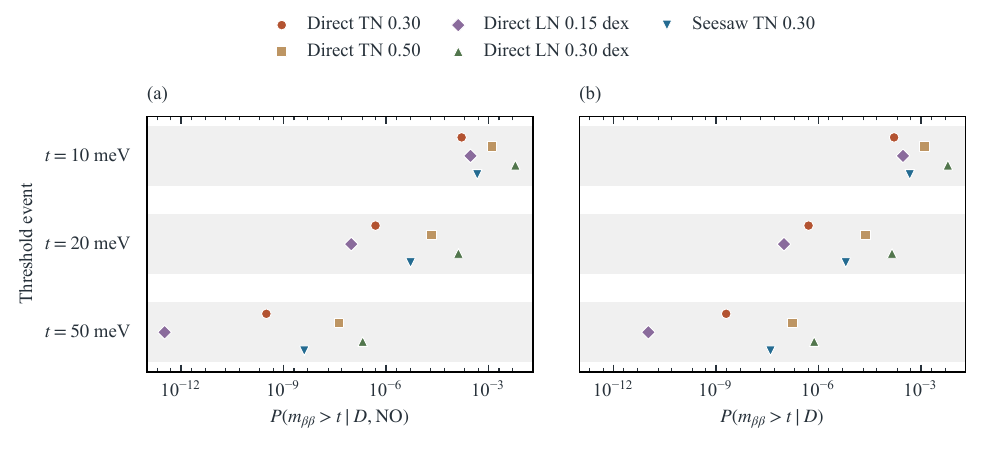}

\caption{Survival probabilities at thresholds $t=10$, 20 and 50 meV: (a) NO; (b) the ordering mixture. Shaded rows are threshold-event categories; vertical marker offsets only separate coefficient measures. Numerical intervals and ordering-conditioned results accompany the data. These are model probabilities rather than detector discovery probabilities.}

\label{fig:C1}

\end{figure*}

The five tail measures comprise direct TN laws with parent width $\sigma_r=0.30$ or 0.50, direct log-normal laws $\log_{10}r\sim N(0,\sigma_{\log}^2)$ with $\sigma_{\log}=0.15$ or 0.30 dex, and the baseline TN seesaw law. Each direct law applies to both charged and Majorana coefficients; all retain uniform primitive phases and the original scale law. They retain a few-meV centre but populate their rare upper regions differently (figure~\ref{fig:C1}). At 20 and 50 meV the largest tested NO probabilities are about $1.3\times10^{-4}$ and $2.1\times10^{-7}$, respectively, while the cross-measure spans grow by several decades. A small IO weight can dominate the surviving far tail \cite{ref31}.

The auxiliary fixed-heavy-mass comparison isolates how observed electron mixing suppresses large contributions. Its reactor-angle condition removes configurations with a large heaviest-state electron projection, consistent with $A_3=m_3\sin^2\theta_{13}$. This fixed-scale diagnostic is separate from the scale-marginalized survival probabilities. Its full observational ablations and numerical sensitivity are archived.

For an ideal event $E$, the incremental seesaw/direct odds factor is $P_S(E\mid D,o)/P_D(E\mid D,o)$ at a stated ordering condition. At 20 meV, an upper condition produces almost no update in these models. Conditional on NO, an exceedance gives factors 10.6 or 0.0396 when the direct TN or LN law is used, illustrating that its direction depends on the specified measures.

For the second texture under NO, the event $m_{\beta\beta}>5$ meV has an incremental seesaw/direct factor 0.352, with numerical 95\% interval $[0.296,0.417]$. The deep-cancellation event $m_{\beta\beta}<1$ meV gives 0.390 with $[0.357,0.426]$. These are incremental odds factors for the stated ideal events. The complete complementary events and uncertainty intervals are released as data. A detector inference instead folds the coherent prediction with a rate likelihood, nuclear matrix elements and detector response.

\newpage
\bibliography{references}

@article{ref1,
  author = {Agostini, M. and Benato, G. and Detwiler, J. A. and Men{\'e}ndez, J. and Vissani, F.},
  title = {Toward the discovery of matter creation with neutrinoless {$\beta\beta$} decay},
  journal = {Rev. Mod. Phys.},
  volume = {95},
  year = {2023},
  pages = {025002},
  doi = {10.1103/RevModPhys.95.025002},
  eprint = {2202.01787},
  archivePrefix = {arXiv}
}

@article{ref2,
  author = {Vissani, F.},
  title = {Signal of neutrinoless double beta decay, neutrino spectrum and oscillation scenarios},
  journal = {JHEP},
  volume = {1999},
  number = {06},
  year = {1999},
  pages = {022},
  doi = {10.1088/1126-6708/1999/06/022},
  eprint = {hep-ph/9906525},
  archivePrefix = {arXiv}
}

@article{ref3,
  author = {Froggatt, C. D. and Nielsen, H. B.},
  title = {Hierarchy of Quark Masses, {Cabibbo} Angles and {CP} Violation},
  journal = {Nucl. Phys. B},
  volume = {147},
  year = {1979},
  pages = {277--298},
  doi = {10.1016/0550-3213(79)90316-X}
}

@article{ref4,
  author = {Feruglio, F. and Romanino, A.},
  title = {Lepton flavor symmetries},
  journal = {Rev. Mod. Phys.},
  volume = {93},
  year = {2021},
  pages = {015007},
  doi = {10.1103/RevModPhys.93.015007},
  eprint = {1912.06028},
  archivePrefix = {arXiv}
}

@article{ref5,
  author = {Hall, L. J. and Murayama, H. and Weiner, N.},
  title = {Neutrino Mass Anarchy},
  journal = {Phys. Rev. Lett.},
  volume = {84},
  year = {2000},
  pages = {2572--2575},
  doi = {10.1103/PhysRevLett.84.2572},
  eprint = {hep-ph/9911341},
  archivePrefix = {arXiv}
}

@article{ref6,
  author = {Haba, N. and Murayama, H.},
  title = {Anarchy and hierarchy: An approach to study models of fermion masses and mixings},
  journal = {Phys. Rev. D},
  volume = {63},
  year = {2001},
  pages = {053010},
  doi = {10.1103/PhysRevD.63.053010},
  eprint = {hep-ph/0009174},
  archivePrefix = {arXiv}
}

@article{ref7,
  author = {Altarelli, G. and Feruglio, F. and Masina, I. and Merlo, L.},
  title = {Repressing Anarchy in Neutrino Mass Textures},
  journal = {JHEP},
  volume = {2012},
  number = {11},
  year = {2012},
  pages = {139},
  doi = {10.1007/JHEP11(2012)139},
  eprint = {1207.0587},
  archivePrefix = {arXiv}
}

@article{ref8,
  author = {Bergstr{\"o}m, J. and Meloni, D. and Merlo, L.},
  title = {{Bayesian} comparison of {U(1)} lepton flavor models},
  journal = {Phys. Rev. D},
  volume = {89},
  year = {2014},
  pages = {093021},
  doi = {10.1103/PhysRevD.89.093021},
  eprint = {1403.4528},
  archivePrefix = {arXiv}
}

@article{ref9,
  author = {Qiu, Y.-C. and Wang, J.-W. and Yanagida, T. T.},
  title = {Predictions of {$m_{ee}$} and neutrino mass from a consistent {Froggatt--Nielsen} model},
  journal = {Phys. Rev. D},
  volume = {108},
  year = {2023},
  pages = {115021},
  doi = {10.1103/PhysRevD.108.115021},
  eprint = {2307.16470},
  archivePrefix = {arXiv}
}

@article{ref10,
  author = {Buchm{\"u}ller, W. and Domcke, V. and Schmitz, K.},
  title = {Predicting {$\theta_{13}$} and the Neutrino Mass Scale from Quark Lepton Mass Hierarchies},
  journal = {JHEP},
  volume = {2012},
  number = {03},
  year = {2012},
  pages = {008},
  doi = {10.1007/JHEP03(2012)008},
  eprint = {1111.3872v3},
  archivePrefix = {arXiv}
}

@article{ref11,
  author = {Ibe, M. and Shirai, S. and Watanabe, K.},
  title = {Comprehensive {Bayesian} Exploration of {Froggatt-Nielsen} Mechanism},
  journal = {JHEP},
  volume = {2025},
  number = {03},
  year = {2025},
  pages = {150},
  doi = {10.1007/JHEP03(2025)150},
  eprint = {2412.19484v2},
  archivePrefix = {arXiv}
}

@article{ref12,
  author = {Cornella, C. and Curtin, D. and Krnjaic, G. and Mellors, M.},
  title = {Testing the {Froggatt-Nielsen} Mechanism with Lepton Flavor and Number Violating Processes},
  journal = {Phys. Rev. D},
  volume = {112},
  year = {2025},
  pages = {115010},
  doi = {10.1103/c2ws-hx4h},
  eprint = {2501.00629v2},
  archivePrefix = {arXiv}
}

@misc{ref13,
  author = {{LEGEND Collaboration}},
  title = {{LEGEND-1000}},
  howpublished = {Official project page},
  url = {https://legend-exp.org/science/legend-pathway/legend-1000},
  note = {Accessed 1 November 2026}
}

@techreport{ref14,
  author = {Abgrall, N. and others},
  collaboration = {LEGEND},
  title = {{LEGEND-1000} Preconceptual Design Report},
  institution = {LEGEND Collaboration},
  year = {2021},
  doi = {10.2172/1811852},
  eprint = {2107.11462},
  archivePrefix = {arXiv}
}

@article{ref15,
  author = {Weinberg, S.},
  title = {Baryon- and Lepton-Nonconserving Processes},
  journal = {Phys. Rev. Lett.},
  volume = {43},
  year = {1979},
  pages = {1566--1570},
  doi = {10.1103/PhysRevLett.43.1566}
}

@article{ref16,
  author = {Minkowski, P.},
  title = {{$\mu\to e\gamma$} at a Rate of One Out of {$10^{9}$} Muon Decays?},
  journal = {Phys. Lett. B},
  volume = {67},
  year = {1977},
  pages = {421--428},
  doi = {10.1016/0370-2693(77)90435-X}
}

@inproceedings{ref17,
  author = {Yanagida, T.},
  title = {Horizontal Gauge Symmetry and Masses of Neutrinos},
  booktitle = {Proceedings of the Workshop on the Unified Theory and the Baryon Number in the Universe},
  editor = {Sawada, O. and Sugamoto, A.},
  publisher = {KEK},
  address = {Tsukuba},
  year = {1979},
  pages = {95},
  reportNumber = {KEK-79-18},
  note = {{KEK-79-18}},
  url = {https://neutrino.kek.jp/seesaw/KEK-79-18-Yanagida.pdf}
}

@article{ref18,
  author = {Lu, X. and Murayama, H.},
  title = {Neutrino Mass Anarchy and the Universe},
  journal = {JHEP},
  volume = {2014},
  number = {08},
  year = {2014},
  pages = {101},
  doi = {10.1007/JHEP08(2014)101},
  eprint = {1405.0547},
  archivePrefix = {arXiv}
}

@article{ref19,
  author = {Huang, G.-y. and Zhou, S.},
  title = {Precise Values of Running Quark and Lepton Masses in the {Standard Model}},
  journal = {Phys. Rev. D},
  volume = {103},
  year = {2021},
  pages = {016010},
  doi = {10.1103/PhysRevD.103.016010},
  eprint = {2009.04851},
  archivePrefix = {arXiv}
}

@article{ref20,
  author = {Esteban, I. and Gonzalez-Garcia, M. C. and Maltoni, M. and Martinez-Soler, I. and Pinheiro, J. P. and Schwetz, T.},
  title = {{NuFit-6.0}: Updated global analysis of three-flavor neutrino oscillations},
  journal = {JHEP},
  volume = {2024},
  number = {12},
  year = {2024},
  pages = {216},
  doi = {10.1007/JHEP12(2024)216},
  eprint = {2410.05380},
  archivePrefix = {arXiv}
}

@misc{ref21,
  author = {Esteban, I. and Gonzalez-Garcia, M. C. and Maltoni, M. and Martinez-Soler, I. and Pinheiro, J. P. and Schwetz, T.},
  title = {{NuFIT 6.1}: Three-neutrino fit based on data available in November 2025},
  year = {2025},
  howpublished = {Numerical release and release notes},
  url = {https://www.nu-fit.org/sites/default/files/v61.release-notes.pdf},
  note = {The analysis uses locally archived {TBoff} and {TByes} grids}
}

@article{ref22,
  author = {Antusch, S. and Drees, M. and Kersten, J. and Lindner, M. and Ratz, M.},
  title = {Neutrino Mass Operator Renormalization Revisited},
  journal = {Phys. Lett. B},
  volume = {519},
  year = {2001},
  pages = {238--242},
  doi = {10.1016/S0370-2693(01)01127-3},
  eprint = {hep-ph/0108005},
  archivePrefix = {arXiv}
}

@article{ref23,
  author = {Varin, C. and Reid, N. and Firth, D.},
  title = {An Overview of Composite Likelihood Methods},
  journal = {Statistica Sinica},
  volume = {21},
  year = {2011},
  pages = {5--42},
  url = {https://www3.stat.sinica.edu.tw/statistica/oldpdf/A21n11.pdf}
}

@article{ref24,
  author = {Haba, N. and Shimizu, Y. and Yamada, T.},
  title = {Neutrino mass square ratio and neutrinoless double-beta decay in random neutrino mass matrices},
  journal = {PTEP},
  volume = {2023},
  year = {2023},
  pages = {023B07},
  doi = {10.1093/ptep/ptad010},
  eprint = {2206.09720},
  archivePrefix = {arXiv}
}

@article{ref25,
  author = {Benato, G.},
  title = {Effective {Majorana} mass and neutrinoless double beta decay},
  journal = {Eur. Phys. J. C},
  volume = {75},
  year = {2015},
  pages = {563},
  doi = {10.1140/epjc/s10052-015-3802-1},
  eprint = {1510.01089},
  archivePrefix = {arXiv}
}

@article{ref26,
  author = {Agostini, M. and Benato, G. and Detwiler, J. A.},
  title = {Discovery probability of next-generation neutrinoless double-{$\beta$} decay experiments},
  journal = {Phys. Rev. D},
  volume = {96},
  year = {2017},
  pages = {053001},
  doi = {10.1103/PhysRevD.96.053001},
  eprint = {1705.02996},
  archivePrefix = {arXiv}
}

@article{ref27,
  author = {Engel, J. and Men{\'e}ndez, J.},
  title = {Status and Future of Nuclear Matrix Elements for Neutrinoless Double-Beta Decay: A Review},
  journal = {Rep. Prog. Phys.},
  volume = {80},
  year = {2017},
  pages = {046301},
  doi = {10.1088/1361-6633/aa5bc5},
  eprint = {1610.06548},
  archivePrefix = {arXiv}
}

@article{ref28,
  author = {Del Moral, P. and Doucet, A. and Jasra, A.},
  title = {Sequential {Monte Carlo} Samplers},
  journal = {J. R. Stat. Soc. B},
  volume = {68},
  year = {2006},
  pages = {411--436},
  doi = {10.1111/j.1467-9868.2006.00553.x}
}

@article{ref29,
  author = {ter Braak, C. J. F.},
  title = {A {Markov Chain Monte Carlo} version of the genetic algorithm {Differential Evolution}: easy {Bayesian} computing for real parameter spaces},
  journal = {Statistics and Computing},
  volume = {16},
  year = {2006},
  pages = {239--249},
  doi = {10.1007/s11222-006-8769-1}
}

@article{ref30,
  author = {Owen, A. B.},
  title = {{Monte Carlo} Variance of Scrambled Net Quadrature},
  journal = {SIAM J. Numer. Anal.},
  volume = {34},
  year = {1997},
  pages = {1884--1910},
  doi = {10.1137/S0036142994277468}
}

@article{ref31,
  author = {Gariazzo, S. and Archidiacono, M. and de Salas, P. F. and Mena, O. and Ternes, C. A. and T{\'o}rtola, M.},
  title = {Neutrino masses and their ordering: Global Data, Priors and Models},
  journal = {JCAP},
  volume = {2018},
  number = {03},
  year = {2018},
  pages = {011},
  doi = {10.1088/1475-7516/2018/03/011},
  eprint = {1801.04946},
  archivePrefix = {arXiv}
}

@article{ref32,
  author = {Antusch, S. and Kersten, J. and Lindner, M. and Ratz, M. and Schmidt, M. A.},
  title = {Running Neutrino Mass Parameters in See-Saw Scenarios},
  journal = {JHEP},
  volume = {2005},
  number = {03},
  year = {2005},
  pages = {024},
  doi = {10.1088/1126-6708/2005/03/024},
  eprint = {hep-ph/0501272},
  archivePrefix = {arXiv}
}

\end{document}